\documentclass[a4paper,11pt]{article}
\usepackage{pos}
\usepackage{amsmath}

\title{Cluster and anti-cluster production in heavy-ion collisions and pA reactions }

\author[a]{Gabriele Coci}
\author[b,c]{Jiaxing Zhao}
\author[d]{Susanne Gl{\"a}ssel}
\author[e]{Viktar Kireyeu}
\author[e]{Vadim Voronyuk}
\author[f]{Michael Winn}
\author[f,g]{J\"org Aichelin}
\author[d,h,b]{Christoph Blume}
\author*[h,b,c]{Elena Bratkovskaya}

\affiliation[a]{Dipartimento di Fisica e Astronomia ``E. Majorana'', Università degli Studi di Catania, Via S. Sofia, 64, I-95125 Catania, Italy}
\affiliation[b]{Helmholtz Research Academy Hessen for FAIR (HFHF), GSI Helmholtz	Center for Heavy Ion Physics. Campus Frankfurt, 60438 Frankfurt, Germany}
\affiliation[c]{Institute for Theoretical Physics, Johann Wolfgang Goethe University, Max-von-Laue-Str. 1, 60438 Frankfurt am Main, Germany }
\affiliation[d]{Institut f\"ur Kernphysik, Max-von-Laue-Str. 1, 60438 Frankfurt, Germany}
\affiliation[e]{ Joint Institute for Nuclear Research, Joliot-Curie 6, 141980 Dubna, Moscow region, Russia}
\affiliation[f]{SUBATECH, Nantes University, IMT Atlantique, IN2P3/CNRS
4 rue Alfred Kastler, 44307 Nantes cedex 3, France}
\affiliation[g]{ Frankfurt Institute for Advanced Studies,
  Ruth Moufang Str. 1, 60438 Frankfurt, Germany}
\affiliation[h]{GSI Helmholtzzentrum für Schwerionenforschung GmbH, Planckstr. 1, 64291 Darmstadt, Germany}

\abstract{We investigate light cluster and anti-cluster production in heavy-ion collisions from SIS to RHIC energies  within the Parton-Hadron-Quantum-Molecular Dynamics (PHQMD)  microscopic transport approach which propagates (anti-)baryons using n-body QMD dynamics. In PHQMD the clusters are formed dynamically by potential interactions between baryons - and recognized by the Minimum Spanning Tree (MST) algorithm - as well as by kinetic reactions in case of deuterons. We present the novel PHQMD results for different observables such as  excitation functions of the multiplicity of deuterons, anti-deuterons and tritons, as well as their transverse momentum spectra. Moreover, we investigate the system size dependence of proton and deuteron production in p+A collisions and show the PHQMD results for p+A collisions (A = Be, Al, Cu, Au) at  14 AGeV/c, as well as for asymmetric Au+A collisions (A =  Al, Cu, Pb) at a bombarding energy of about 10 AGeV.
}

\FullConference{10th Int. Conference on ”Quarks and Nuclear Physics” (QNP-2024)\\
University of Barcelona, Spain, 8-12 July, 2024}

\begin{document}
\maketitle

\section{Introduction}

The observation of light baryonic clusters and anti-clusters in the central rapidity region originated from ultra-relativistic heavy-ion collisions has gained significant interest and research activity in recent years, both theoretically and experimentally. A key physics question that has emerged is how such weakly bound objects can form and persist within the hot, dense environment created in the central collision zone. The identification of mechanisms of cluster production via the comparison of experimental data with theoretical calculations is a challenging methodological task. In this respect the modeling of cluster production within realistic dynamical approaches, as well as their comparison to sensitive observables, are of special interest. 

In this proceedings we report on  the novel PHQMD results on the excitation functions and spectra of light clusters and of anti-clusters as well as on the system size dependence of deuteron production in p+A reactions and for asymmetric Au+A collisions at tens GeV energies.  The PHQMD calculations  are confronted with the existing experimental data.

\section{Cluster and anti-cluster production within PHQMD}

Our study is based on the  Parton-Hadron-Quantum-Molecular Dynamics (PHQMD)  microscopic transport approach   \cite{Aichelin:2019tnk,Glassel:2021rod,Kireyeu:2022qmv,Coci:2023daq,Kireyeu:2023bye}. PHQMD is a microscopic n-body transport model based on the QMD propagation of the baryonic degrees-of-freedom and the dynamical properties and interactions in- and out-of-equilibrium of hadronic and partonic degrees-of-freedom of the Parton-Hadron-String-Dynamics (PHSD) approach \cite{Cassing:2009vt,Linnyk:2015rco}.

Within PHQMD the clusters are formed dynamically, via a 'potential' mechanism, i.e. by potential interactions between nucleons and hyperons (and the anti-baryons), and recognized by  the advanced Minimum Spanning Tree (aMST) algorithm,  which is identifying bound clusters (with negative binding energies) by correlations of baryons in coordinate space \cite{Coci:2023daq}. 
Additionally,  the 'kinetic' mechanism for deuteron production is incorporated  by catalytic hadronic reactions accounting for all isospin channels of the various $\pi NN\leftrightarrow \pi d$, $NNN\leftrightarrow N d$ reactions. Furthermore, we account for the quantum nature of the deuteron by means of its finite size, which is modeled by a finite-size excluded volume in coordinate space and the projection of the relative momentum of the interacting pair of nucleons on the deuteron wave-function in momentum space. These quantum effects lead to a strong reduction of $d$ production by kinetic reactions, especially at target/projectile rapidities \cite{Coci:2023daq}.
The production of anti-clusters (build from anti-baryons) occurs by potential interactions of anti-baryons and is recognized by the aMST in a similar way as for clusters. We note that in this study we use a hard static potential which leads to a hard equation-of-state  \cite{Aichelin:2019tnk}.

\section{Results }

\begin{figure}[h!]
    \centering
 \resizebox{1.\textwidth}{!}{
\includegraphics[scale=1]{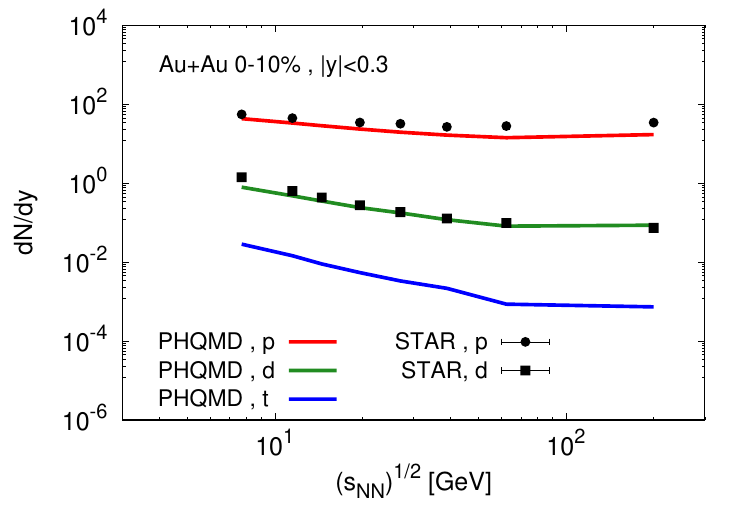 }
\hspace*{-7mm}
\includegraphics[scale=1]{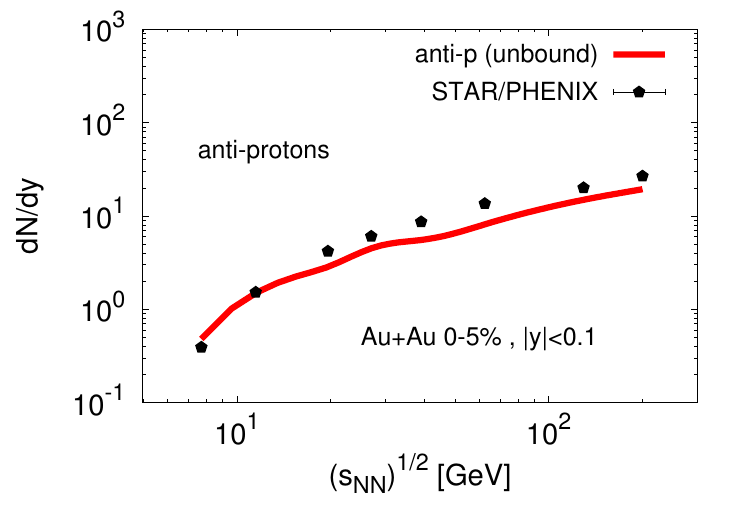 }
\hspace*{-7mm}
\includegraphics[scale=1]{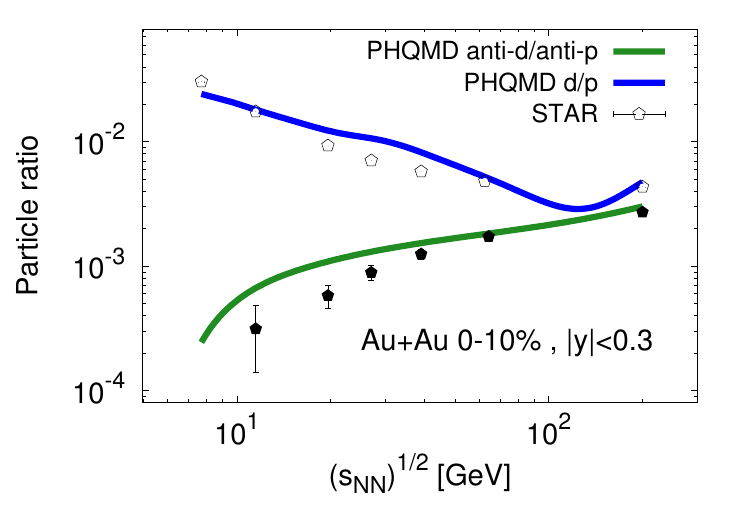 }}
\caption{ 
The PHQMD results for the excitation function for $dN/dy$ of protons (red line), deuterons (green line), tritons (blue line) - left plot, anti-protons - middle plot - and ratios of $d/p$ (green line) and $\bar d/\bar p$ (blue line) - right plot-  as a function of $\sqrt{s_{NN}}$ for  $0-10\%$ central Au+Au collisions at midrapidity $|y|<0.3$ in comparison with the experimental data from the STAR collaboration~\cite{STAR:2019sjh}. }    
    \label{Fig:antid}     
\end{figure}

In Fig. \ref{Fig:antid}  we present the PHQMD results for the excitation function of the midrapidity  $dN/dy$ of free (unbound) protons, deuterons, tritons (left plot), anti-protons (middle plot) and ratios of $d/p$ and $\bar d/\bar p$ (right plot) as a function of $\sqrt{s_{NN}}$ for  $0-10\%$ central Au+Au collisions in comparison with the experimental data from the STAR collaboration~\cite{STAR:2019sjh}.  As follows from Fig. \ref{Fig:antid} the PHQMD reproduces the experimental data quite well for the whole energy range from $\sqrt{s_{NN}} = 7.7$ GeV to 200 GeV and also in line with the UrQMD calculations \cite{Sombun:2018yqh} combined with a coalescence model.
One can see from the left plot that the number of protons, as well as deuterons and tritons, at midrapidity decreases with increasing bombarding energy, while the number of anti-protons (middle plot) increases. We note that in the PHSD/PHQMD approach the multi-meson fusion reactions ($B +\bar B \Leftrightarrow  3 \ mesons$)  are included by default \cite{Cassing:2001ds} which partially compensates the anti-baryon absorption \cite{Seifert:2017oyb}.
This also leads to the increase of the $\bar d/\bar p$ ratio (right plot) while the ratio $d/p$ decreases with energy.

\begin{figure}[h!]
    \centering
\includegraphics[scale=0.5]{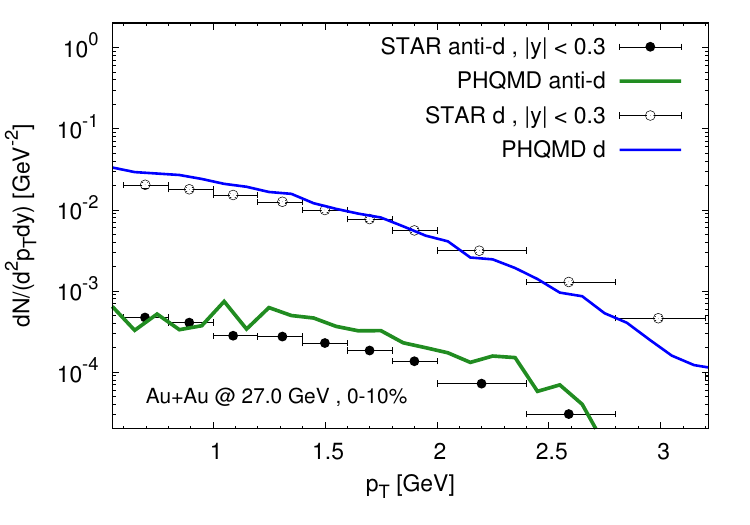}
\includegraphics[scale=0.5]{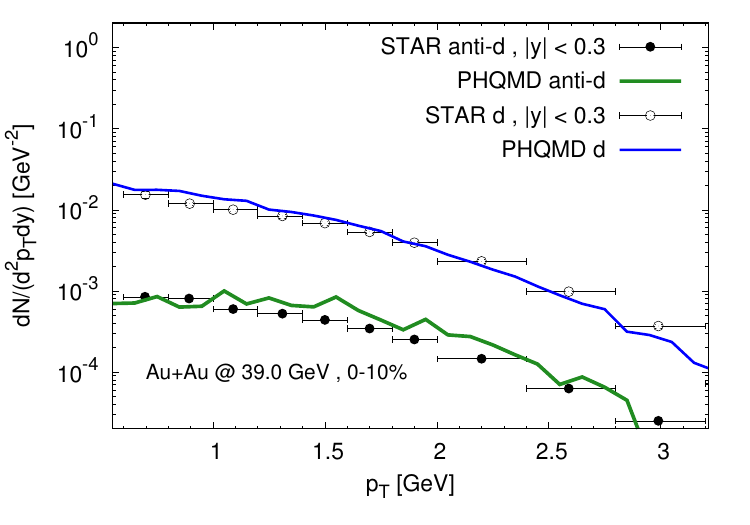} \\
\includegraphics[scale=0.5]{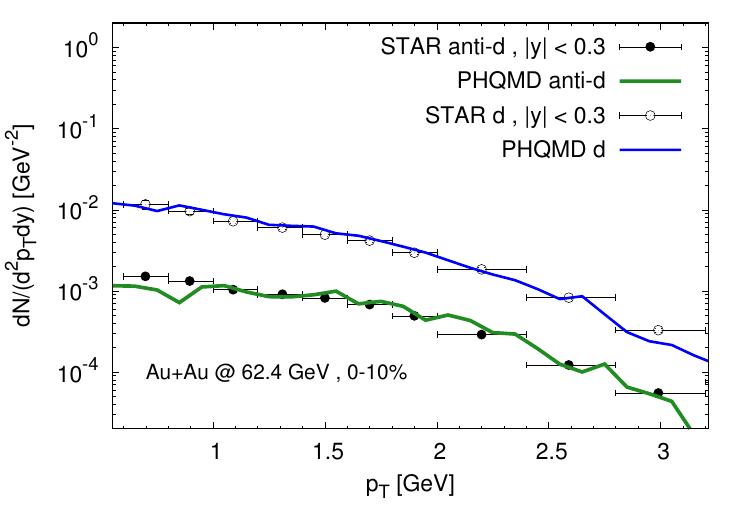}
\includegraphics[scale=0.5]{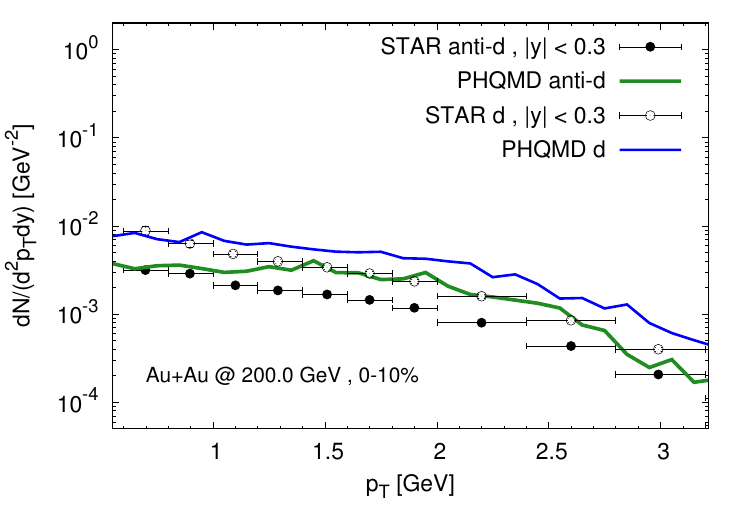}
\caption{
The PHQMD results for transverse momentum distributions of deuterons (blue lines) and anti-deuterons (green lines) from Au+Au central collisions at RHIC BES energies $\sqrt{s_{NN}}=$ 27, \ 39, \ 62.4, \ 200 GeV in comparison to the experimental data from the STAR collaboration~\cite{STAR:2019sjh}.}    
    \label{Fig:ptRHIC}     
\end{figure}

The PHQMD results for transverse momentum distributions of deuterons (blue lines) and anti-deuterons (green lines) from Au+Au central collisions at the RHIC BES energies $\sqrt{s_{NN}}=$ 27, \ 39, \ 62.4, \ 200 GeV are shown in Fig. \ref{Fig:ptRHIC} in comparison with the experimental data from the STAR collaboration~\cite{STAR:2019sjh}. One can see that PHQMD follows the experimental observations within the error bars and reproduces approximately the slopes for deuterons and anti-deuterons.
As has been shown in Ref.  \cite{Coci:2023daq}, within PHQMD  the potential mechanism is dominant for the deuteron production at all relativistic energies. Similar, the anti-deuterons are formed from comoving anti-protons and anti-neutrons and recognized by  aMST.

\begin{figure}[h!]
\centering
\includegraphics[width=1.0\textwidth]{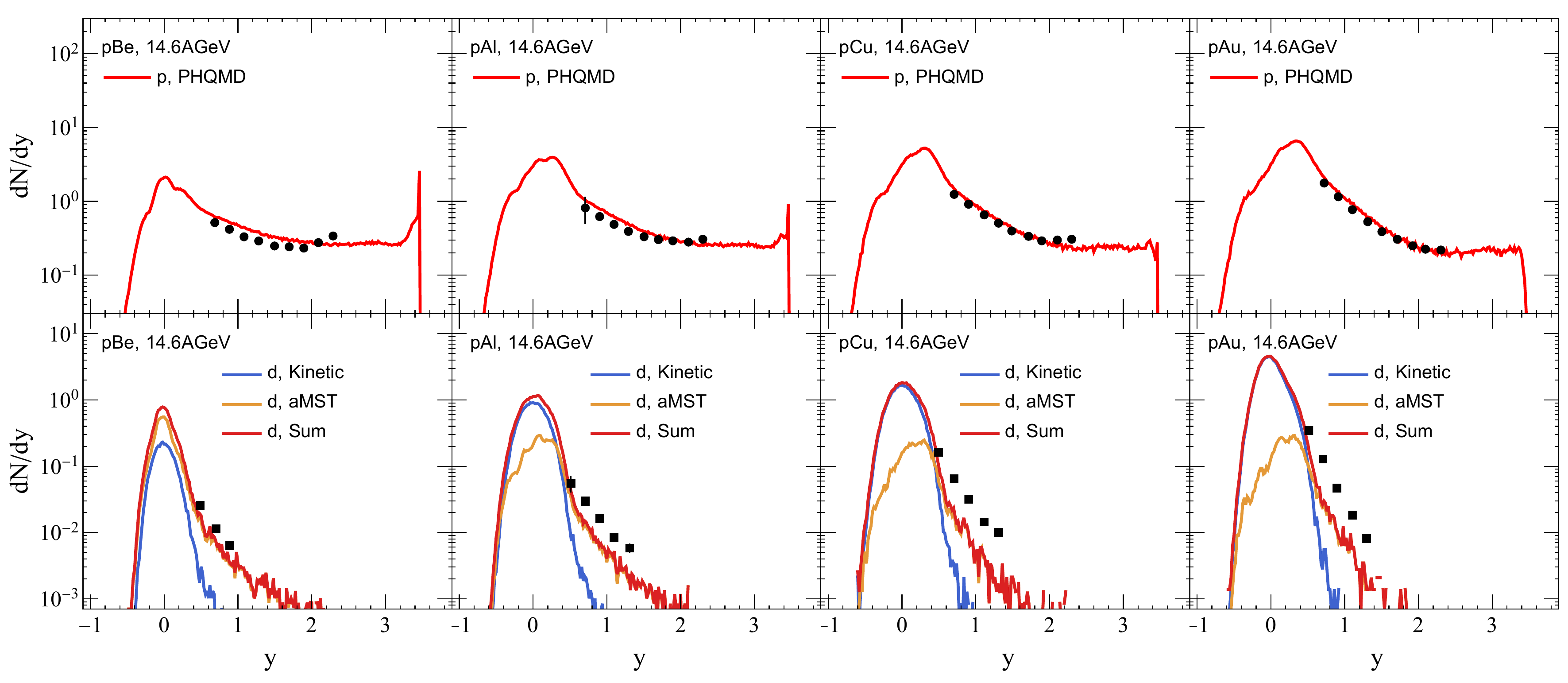}
\caption{
The PHQMD results for the proton (upper row) and deuteron (lower row) rapidity distributions from p+Be, p+Al, p+Cu, and p+Au collisions at $14.6 ~\rm AGeV/c$. The blue lines in the lower row indicate the kinetic deuterons, the orange lines - the potential deuterons with negative binding energies identified by the aMST, the red lines show the sum of both production mechanisms.
The experimental data are from the E802 Collaboration at the BNL-AGS~\cite{E-802:1991unu}.  }   
\label{Fig:ypA}     
\end{figure}

Light clusters can be formed not only in A+A collisions, but also in  proton-induced (p+A) reactions on nuclei of different size $A$. Such  measurements impose challenging constraints on the theoretical models, since it allows to access the cluster production of (near) target rapidities where heavy-ion data are not available at relativistic energies so far.
In Fig. \ref{Fig:ypA} we present the PHQMD results for the free proton (upper row) and deuteron (lower row) rapidity distributions in p+Be, p+Al, p+Cu, and p+Au collisions at 14.6  AGeV/c in comparison with the  experimental data from the E802 Collaboration at the BNL-AGS~\cite{E-802:1991unu}.  
As follows from the upper panel, PHQMD reproduces the proton stopping very well for all systems. The  deuteron rapidity distribution (shown on the lower panel) is very well reproduced for the light Be target and slightly underestimated for more heavy targets. Moreover, the kinetic mechanism is very important for the target region where the nucleon density is larger than at midrapidity where the MST finds more deuterons bound by potential interactions.

\begin{figure}[h!]
\centering
\includegraphics[width=1.0\textwidth]{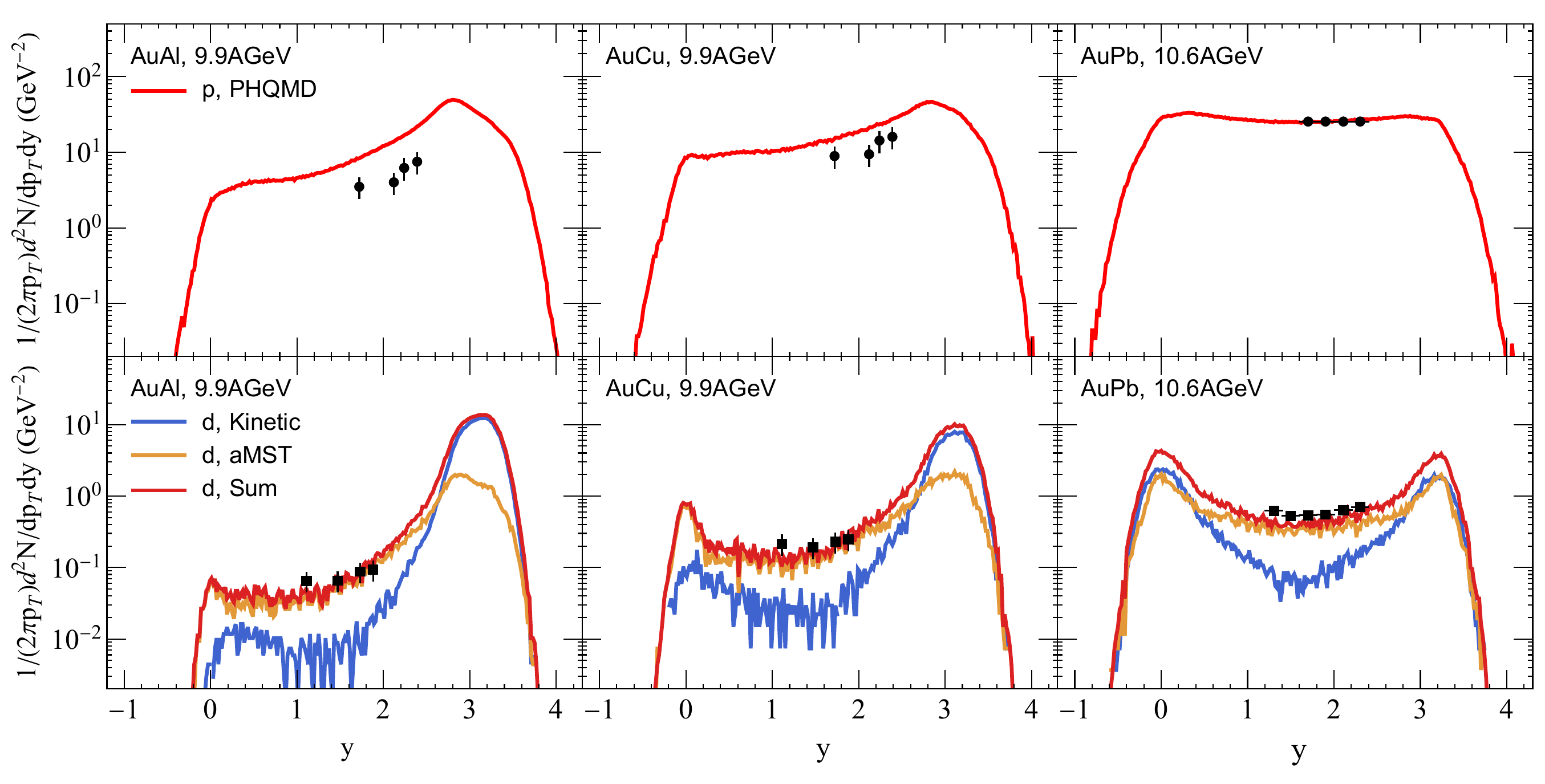}
\caption{
The PHQMD results for the proton (upper row) and deuteron (lower row) rapidity distributions at small $p_T$  ($p_T<0.03$ GeV$/c$ in the PHQMD calculations) from Au+Al and Au+Cu  collisions at $E_{beam}=9.9$ AGeV and from Au+Pb collisions at 10.6 AGeV. The blue lines in the lower row indicate the kinetic deuterons, the orange lines - the potential deuterons with negative binding energies identified by the aMST, the red lines show the sum of both production mechanisms.
The experimental data for Au+Al and Au+Cu are from the E878 Collaboration \cite{E878:1998vna} and for Au+Pb from the E864 Collaboration \cite{E864:2000auv}, both at the BNL-AGS.  }   
\label{Fig:yAuA}     
\end{figure}

Finally, we step to the deuteron production in the asymmetric Au+A systems.
In Fig.\ref{Fig:yAuA} we show the PHQMD results for the proton  and deuteron  rapidity distributions from  Au+Al and Au+Cu  collisions at $E_{beam}=9.9$ AGeV and from Au+Pb collisions at 10.6 AGeV in comparison with the experimental data from the E878 Collaboration \cite{E878:1998vna} and for Au+Pb from the E864 Collaboration \cite{E864:2000auv}.  One can see that the proton data for Au+Al and Au+Cu from the E878 Collaboration is overestimated by PHQMD, while it is in a very good agreement with the E864 data for Au+Pb collisions. This can be attributed to a poor acceptance of E878, as discussed in Ref.~\cite{E864:2000auv}, which is not applied to the PHQMD calculations.
The deuteron spectra are well reproduced for all systems. One can see that the dominant mechanism for the deuteron production at midrapidity is the potential one similar to the symmetric A+A collisions.

\section{Summary}
We presented the PHQMD results for the excitation function of the proton, deuteron, triton as well as for anti-proton and anti-deuteron multiplicities, in the wide energy range of  $\sqrt{s_{NN}} =7.7 -200$ AGeV, as well as their $p_T$ spectra.  
Moreover, we estimated the system size dependence of the  proton stopping and the deutron production for proton-induced (p+A) and asymmetric Au+A collisions at AGS energies.
We show that the PHQMD provides a reasonably good description of experimental data for protons and light clusters $(d, t)$, as well as for anti-protons and anti-deuterons for all studied observables and systems. We find that at midrapidity the deutrons are dominantly produced by the potential interactions for Au+Au at all studied energies as well as for p+A and Au+A collisions at AGS energies.



\bibliographystyle{abbrv}
\bibliography{references.bib}

\begin{thebibliography}{10}

\bibitem{E-802:1991unu}
T.~Abbott et~al.
\newblock {Measurement of particle production in proton induced reactions at
  14.6-GeV/c}.
\newblock {\em Phys. Rev. D}, 45:3906--3920, 1992.

\bibitem{STAR:2019sjh}
J.~Adam et~al.
\newblock {Beam energy dependence of (anti-)deuteron production in Au + Au
  collisions at the BNL Relativistic Heavy Ion Collider}.
\newblock {\em Phys. Rev. C}, 99(6):064905, 2019.

\bibitem{Aichelin:2019tnk}
J.~Aichelin, E.~Bratkovskaya, A.~Le~F\`evre, V.~Kireyeu, V.~Kolesnikov,
  Y.~Leifels, V.~Voronyuk, and G.~Coci.
\newblock {Parton-hadron-quantum-molecular dynamics: A novel microscopic $n$
  -body transport approach for heavy-ion collisions, dynamical cluster
  formation, and hypernuclei production}.
\newblock {\em Phys. Rev. C}, 101(4):044905, 2020.

\bibitem{E864:2000auv}
T.~A. Armstrong et~al.
\newblock {Measurements of light nuclei production in 11.5-A-GeV/c Au + Pb
  heavy ion collisions}.
\newblock {\em Phys. Rev. C}, 61:064908, 2000.

\bibitem{E878:1998vna}
M.~J. Bennett et~al.
\newblock {Light nuclei production in relativistic Au + nucleus collisions}.
\newblock {\em Phys. Rev. C}, 58:1155--1164, 1998.

\bibitem{Cassing:2001ds}
W.~Cassing.
\newblock {Anti-baryon production in hot and dense nuclear matter}.
\newblock {\em Nucl. Phys. A}, 700:618--646, 2002.

\bibitem{Cassing:2009vt}
W.~Cassing and E.~L. Bratkovskaya.
\newblock {Parton-Hadron-String Dynamics: an off-shell transport approach for
  relativistic energies}.
\newblock {\em Nucl. Phys. A}, 831:215--242, 2009.

\bibitem{Coci:2023daq}
G.~Coci, S.~Gl\"a\ss{}el, V.~Kireyeu, J.~Aichelin, C.~Blume, E.~Bratkovskaya,
  V.~Kolesnikov, and V.~Voronyuk.
\newblock {Dynamical mechanisms for deuteron production at mid-rapidity in
  relativistic heavy-ion collisions from energies available at the GSI
  Schwerionensynchrotron to those at the BNL Relativistic Heavy Ion Collider}.
\newblock {\em Phys. Rev. C}, 108(1):014902, 2023.

\bibitem{Glassel:2021rod}
S.~Gl\"a\ss{}el, V.~Kireyeu, V.~Voronyuk, J.~Aichelin, C.~Blume,
  E.~Bratkovskaya, G.~Coci, V.~Kolesnikov, and M.~Winn.
\newblock {Cluster and hypercluster production in relativistic heavy-ion
  collisions within the parton-hadron-quantum-molecular-dynamics approach}.
\newblock {\em Phys. Rev. C}, 105(1):014908, 2022.

\bibitem{Kireyeu:2023bye}
V.~Kireyeu, G.~Coci, S.~Glaessel, J.~Aichelin, C.~Blume, and E.~Bratkovskaya.
\newblock {Cluster formation near midrapidity -- can the mechanism be
  identified experimentally?}
\newblock 4 2023.

\bibitem{Kireyeu:2022qmv}
V.~Kireyeu, J.~Steinheimer, J.~Aichelin, M.~Bleicher, and E.~Bratkovskaya.
\newblock {Deuteron production in ultrarelativistic heavy-ion collisions: A
  comparison of the coalescence and the minimum spanning tree procedure}.
\newblock {\em Phys. Rev. C}, 105(4):044909, 2022.

\bibitem{Linnyk:2015rco}
O.~Linnyk, E.~L. Bratkovskaya, and W.~Cassing.
\newblock {Effective QCD and transport description of dilepton and photon
  production in heavy-ion collisions and elementary processes}.
\newblock {\em Prog. Part. Nucl. Phys.}, 87:50--115, 2016.

\bibitem{Seifert:2017oyb}
E.~Seifert and W.~Cassing.
\newblock {Baryon-antibaryon annihilation and reproduction in relativistic
  heavy-ion collisions}.
\newblock {\em Phys. Rev. C}, 97(2):024913, 2018.

\bibitem{Sombun:2018yqh}
S.~Sombun, K.~Tomuang, A.~Limphirat, P.~Hillmann, C.~Herold, J.~Steinheimer,
  Y.~Yan, and M.~Bleicher.
\newblock {Deuteron production from phase-space coalescence in the UrQMD
  approach}.
\newblock {\em Phys. Rev. C}, 99(1):014901, 2019.

\end{thebibliography}

\end{document}